\documentclass[%
preprint%
 ,secnumarabic%
,amssymb, amsmath,nobibnotes,aps]{revtex4}

\usepackage{epsfig}
\usepackage{graphics}
\usepackage{float}
\expandafter\ifx\csname package@font\endcsname\relax\else
 \expandafter\expandafter
 \expandafter\usepackage
 \expandafter\expandafter
 \expandafter{\csname package@font\endcsname}%
\fi

\begin{document}

\title{ Triplet-Singlet Extension of the MSSM with a 125 GeV Higgs Boson and Dark Matter}

\author{Tanushree Basak}
\email{tanu@prl.res.in}
\author{Subhendra Mohanty}
\email{mohanty@prl.res.in}
\affiliation{Physical Research Laboratory, Ahmedabad 380009,
India.}
\def\be{\begin{equation}}
\def\ee{\end{equation}}
\def\al{\alpha}
\def\bea{\begin{eqnarray}}
\def\eea{\end{eqnarray}}

\begin{abstract}

We study the extension of Minimal Supersymmetric Standard Model by adding one singlet and one hypercharge zero SU(2) triplet chiral
superfield. The triplet sector gives an additional contributions to the scalar masses and we find that the lightest
CP-even Higgs boson can have a mass of 119-120 GeV at the tree level, and radiative correction raises the value to 125 GeV. 
In this model no significant contributions from stop loops is needed to get the required Higgs mass which
alleviates the fine tuning problem of fixing the stop mass to a high precision at the GUT scale. In addition this model gives a
neutralino dark matter of mass around 100 GeV which is a mixture of Higgsino and Triplino with a dark matter density
 consistent with WMAP observations.
The spin-independent scattering cross-section with nucleons is $10^{-43} cm^2$, which makes it consistent with the
bounds from direct detection experiments like XENON100 and others.

\end{abstract}
\maketitle
\section{Introduction}

The ATLAS  and CMS  collaborations \cite{ATLAS,CMS} have narrowed down the allowed range of a
light Higgs mass to the region 115-131 GeV. In addition there are hints of the Higgs mass being
near $m_h=125$GeV with SM like decay widths into $2\gamma$ and $4l$. A light Higgs is favored
in supersymmetry although the MSSM predicts a tree level upper bound on the
lightest CP-even Higgs mass as $m_h < M_Z \cos{2\beta}$. Within MSSM, loop corrections can
give required large corrections to Higgs mass provided the stop is heavier than 1 TeV or there is 
near maximal stop mixing.  Implications  of the 125 GeV Higgs for the MSSM and constrained-MSSM
parameter space have been extensively studied \cite{MSSM}. Going beyond MSSM, in order to get a larger
tree-level Higgs mass, the simplest extension is a singlet superfield in the NMSSM model \cite{NMSSM}.
The singlet interaction with the two Higgs doublet of MSSM is via the $\lambda S H_u\cdot H_d$ term.
The  Higgs mass is now given by the relation
$m_h^2 = M_Z^2 \cos^2 2\beta + \lambda^2 v^2 \sin^2 2\beta+ \delta m_h^2$, where $\delta m_h^2 $
is due to radiative-correction. Taking $\lambda=0.7$ (larger values would make it flow to
the non-perturbative regime much below the GUT scale) and $\tan \beta=2$ the
radiative correction needed to get a 125 GeV Higgs mass is $\delta m_h=55$GeV which is an improvement
over the $\delta m_h=85$GeV needed in the MSSM. However fine-tuning of the stop mass is still
required in NMSSM to get the required Higgs mass \cite{King}. Also by extending the MSSM gauge group in a suitable 
way, the new Higgs sector dynamics can push the tree-level mass well above the tree-level MSSM limit 
if it couples to the new gauge sector \cite{d-term-MSSM}. In most of the cases the non-decoupling D-terms 
contribute non-trivially to increase the tree-level mass of the SM-like Higgs boson. Recent anaysis of the SUSY model 
based on $SU(3)_C \times SU(2)_L \times U(1)_R \times U(1)_{B-L}$ gauge group \cite{Hirsch} has shown that 
the tree level physical Higgs mass can be atmost 110 GeV and through the one-loop correction it can be raised considerably.  
Another recent work on MSSM extended by a U(1) gauged Peccei-Quinn symmetry \cite{an-liu-wang} where the new D-terms 
can raise the tree-level mass well enough to accommodate the 125 GeV Higgs boson without significant radiative correction 
and hence requires less fine-tuning.

An important aspect of the 125 GeV Higgs mass is that the parameter space of thermal relic for
dark matter is severely restricted. In MSSM, the LSP is a Higgsino at the TeV scale \cite{dark-matter-MSSM}.
In NMSSM, SUSY partner of the singlet scalar -the singlino mixes with the neutralinos to provide a light dark
matter \cite{Draper:2010ew,Cao}.
Recent analysis \cite{King} has shown that the benchmark parameters which give a 125 GeV Higgs
also provide a neutralino dark matter candidate with mass in the range of 68-85 GeV.
To our knowledge, the dark matter in triplet-extended MSSM has not been studied so far.

The extension of MSSM by extending it with a $Y=0$ and $Y=0,\pm 1$  SU(2)
 triplet superfields has been studied  \cite{Espinosa:1991gr,Espinosa:1991wt,DiChiara:2008rg}
where  the tree level contribution to the Higgs mass from the triplet Higgs sector has been calculated. It has been shown
in \cite{DiChiara:2008rg} that with the $Y=0$ triplet superfield the tree-level Higgs mass can be raised to 113 GeV which would
still require substantial loop corrections from stops.
Recently, the MSSM extended by two real triplets ($Y=\pm 1$) and one singlet\cite{Agashe:2011ia}
has been studied with a motivation to solve the $\mu$-problem as well as to obtain  a large
correction to the lightest Higgs mass. The analysis of the dark matter sector of this model will be complicated as the
LSP will be the lightest eigenstate of the $7\times 7$ neutralino mass matrix which has not yet been done.

In this paper we explore the minimal extension of the MSSM which can give a tree-level Higgs mass of 119-120 GeV. We find that by
extending the MSSM by adding a singlet and a $Y=0$ SU(2) triplet superfields,  this aim can be achieved.
The upper bound on the tree-level mass of the
lightest CP-even Higgs is given in equation (\ref{bound}). With this tree level Higgs mass the stop mass need not be very heavy and this
solves the fine-tuning problem of the Higgs mass in MSSM and NMSSM \cite{King}. We also study the dark matter candidates in this model which
is obtained by diagonalizing the $6\times 6$ neutralino mass matrix. We find a viable dark matter with mass 100 GeV, which is a mixture
of the Higgsino and Triplino (the fermionic partner of the neutral component of the triplet Higgs). We fix two sets of benchmark parameters
at the electroweak scale which would give a the 125 GeV and dark matter relic density $\Omega h^2= 0.1109 \pm 0.0056$ compatible with
WMAP-7 measurements \cite{Wmap}. We find that the direct detection cross section of the dark matter is $\sigma_{SI} \simeq 10^{-43} cm^2$,
which is compatible with the direct detection experiments like XENON100 \cite{Baudis:2012zs}.

In Section(\ref{model}) we display the superpotential of our model
and  we derive the
scalar potential from the D-terms and F-terms and from the various soft-breaking terms. In Section(\ref{higgs}) we
give a detailed analysis of the Higgs sector and we calculate
the CP-even, CP-odd and Charged Higgs mass matrices. In Section(\ref{neut}), the neutralino and the chargino mass
matrices are discussed. The numerical results based on this model are discussed in detail in Section(\ref{results}).
We show  the results for two sets of benchmark points which include the parameters like couplings, tri-linear soft 
breaking terms, soft masses and the fermionic and scalar mass spectrum. We have also taken into account the one-loop 
corrections to the lightest physical Higgs mass and shown a quantitative improvement of the level of fine-tuning 
compared to other models. In Section(\ref{DM}) we discuss the  Dark Matter from the neutralino sector of this model
and its phenomenology, which is one of the main results of this paper. In the Concluding section we 
summarize the results and we point out some directions for further study of the Triplet-Singlet model 
which will enable the model to be tested at the LHC.

\section{Model}
\label{model}

In this model, we have extended the superpotential of the minimal supersymmetric standard model
by adding one singlet chiral superfield $S$ and one SU(2) triplet chiral superfields $T_{0}$ with hypercharge $Y = 0$.
The most general form of the superpotential for this singlet-triplet extended
model can be written as,
\small{
\begin{eqnarray}
{\mathcal W}&=&(\mu + \lambda \hat{S}) \hat{H_{d}}.\hat{H_{u}}+ \frac{\lambda_1}{3} \hat{S}^3+\lambda_2 \hat{H_{d}}.\hat{T_{0}}\hat{H_{u}}+
\lambda_{3} \hat{S}^2 Tr(\hat{T}_0)+\lambda_4 \hat{S}Tr(\hat{T}_{0}\hat{T}_{0})+W_{Yuk.}
\end{eqnarray}}
where, $\hat{H}_{u,d}$ are the Higgs doublets of the MSSM and the Yukawa superpotential $ W_{Yuk.}$ is given as,
\small{
\begin{equation}
 W_{Yuk.}=y_u\hat{Q}_L.\hat{H}_u\hat{U}_R+y_d\hat{Q}_L.\hat{H}_d\hat{D}_R+y_e\hat{L}_L.\hat{H}_d\hat{E}_R
  \end{equation}}
In terms of the components, we have
\begin{center}
             $\hat{H}_u=\begin{pmatrix}
                   \hat{H}_u^+\\
                    \hat{H}_u^0
                  \end{pmatrix} $,
              $\hat{H}_d=\begin{pmatrix}
                   \hat{H}_d^0\\
                    \hat{H}_d^-
                  \end{pmatrix} $ and
            $ \hat{T}_{0}=\begin{pmatrix}
                \frac{\hat{T}^{0}}{\sqrt{2}} & -\hat{T}^{+}_{0}\\
                \hat{T}^{-}_{0} &  \frac{-\hat{T}^{0}}{\sqrt{2}}
               \end{pmatrix} $
\end{center}
Here, $(\hat{T}^-_0)^*\neq -\hat{T}^{+}_{0}$, which would not have been true for real Higgs triplet
in non-supersymmetric models.
We can solve the $\mu$-problem by starting with a  scale invariant superpotential, given as
\small{
\begin{eqnarray}
W_{sc.inv.}&=&\lambda \hat{S}\hat{H_{d}}.\hat{H_{u}}+ \frac{\lambda_1}{3} \hat{S}^3+\lambda_2 \hat{H_{d}}.\hat{T_{0}}\hat{H_{u}}+
\lambda_4 \hat{S}Tr(\hat{T}_{0}\hat{T}_{0})+W_{Yuk.}
\label{W_s}
\end{eqnarray}}
where the SU(2) invariant dot product is defined as,
\small{
\begin{eqnarray}
\hat{H}_{d}.\hat{T}_{0}\hat{H}_{u}&=& \frac{1}{\sqrt{2}}(\hat{H}_{d}^0\hat{T}^0\hat{H}_{u}^0 + \hat{H}_{d}^{-}\hat{T}^{0}\hat{H}_{u}^{+})
-(\hat{H}_{d}^{0}\hat{T}_{0}^{-}\hat{H}_{u}^{+} + \hat{H}_{d}^{-}\hat{T}_{0}^{+}\hat{H}_{u}^{0})
\end{eqnarray}}
This superpotential(\ref{W_s}) also has an accidental
$Z_3$-symmetry, i.e. invariance of the superpotential on  multiplication of
 the chiral superfields  by the factor of $\frac{2\pi i}{3}$. By, this choice we
are eliminating the $\mu$-parameter but an effective $\mu$-term is generated when the
neutral components of $S$ and $T_0$ acquire vev's $v_s$ and $v_t$ respectively,
\begin{equation}
 \mu_{eff}= \lambda v_s-\frac{\lambda_2}{\sqrt{2}} v_t
\label{mu}
\end{equation}

Therefore, in terms of the neutral components of the super-fields the equation (\ref{W_s}) sans $W_{Yuk.}$ can be re-written as,
\small{
\begin{equation}
 W^{neu}=-\lambda \hat{S} \hat{H}_u^0 \hat{H}_d^0 +\frac{\lambda_1}{3} \hat{S}^3 +\frac{\lambda_2}{\sqrt{2}} \hat{H}_{d}^0\hat{T}^0\hat{H}_{u}^0
+\lambda_{4} \hat{S} \hat{T}^0 \hat{T}^0
\label{W-neu}
\end{equation}}

\subsection{Scalar potential}
\label{scalarpot}
The scalar potential involving only Higgs field can be written as,
\begin{eqnarray}
V&=& V_{SB} + V_{F} + V_{D}
\label{scalarV}
\end{eqnarray}
In the above equation, $V_{SB}$ consists of the soft-supersymmetry breaking term
associated with the superpotential in equation(\ref{W_s}), is given by
\small{
\begin{eqnarray}
 V_{SB}&=& m_{H_{u}}^{2}[\lvert H_{u}^{0}\rvert^{2}+\lvert H_{u}^{+}\rvert^{2}] + m_{H_{d}}^{2}[\lvert H_{d}^{0}\rvert^{2}
+\lvert H_{d}^{-}\rvert^{2}]+m_S^2 \lvert S\rvert^{2}+ m_{T}^{2}Tr(T_0^\dag T_0)+\nonumber \\
&& (-\lambda A_{\lambda} S H_u. H_d +\frac{\lambda_1}{3}A_{\lambda_1} S^3 +\lambda_2 A_{\lambda_2} H_{d}.T_0H_{u}+
\lambda_{4} B_{\lambda}S Tr(T_0^2)+h.c)
\end{eqnarray}
}
In equation(\ref{scalarV}) $V_{F}$ is the supersymmetric potential from F-terms, given by
\small{
\begin{eqnarray}
 V_{F}&=& \lvert -\lambda S H_d^0+\frac{\lambda_2}{\sqrt{2}}H_d^0 T^0 -\lambda_2 H_d^- T_0^+ \rvert^2 +
\lvert -\lambda S H_u^0+\frac{\lambda_2}{\sqrt{2}}H_u^0 T^0 -\lambda_2 H_u^+ T_0^-\rvert^2 \nonumber\\ && +
 \lvert \frac{\lambda_2}{\sqrt{2}}(H_u^0 H_d^0 +H_d^- H_u^+)+2\lambda_4 S T^0 \rvert^2 +
\lvert \lambda (H_d^- H_u^+H_u^0 H_d^0)+\lambda_1 S^2+\lambda_4 (T^{0^2}-2T_0^+ T_0^-)\rvert^2 \nonumber \\ && +
\lvert \lambda S H_d^- +\frac{\lambda_2}{\sqrt{2}}T^0H_d^- - \lambda_2H_d^0 T_0^- \rvert^2 + \lvert -\lambda_2H_d^-H_u^0-2\lambda_4 ST_0^- \rvert^2 \nonumber \\&& +
\lvert \lambda S H_u^+ +\frac{\lambda_2}{\sqrt{2}}T^0H_u^+ - \lambda_2H_u^0 T_0^+\rvert^2 + \lvert -\lambda_2H_u^+H_d^0-2\lambda_4 ST_0^+\rvert^2
\end{eqnarray}
}
whereas the F-term for the neutral scalar potential can be derived from equation(\ref{W-neu}) as,
\small{
\begin{eqnarray}
 V_{F_{neu}} &=& \sum_i \lvert \frac{\partial W^{neu}_{scalar}}{\partial \phi_i^0}\rvert^2
\end{eqnarray}
} where, $\phi_i^0$ stands for $H_u^0, H_d^0, S, T^0$ and $W^{neu}_{scalar}$ is the scalar counter-part of the neutral superpotential $W^{neu}$.

Finally, $V_{D}$ is supersymmetric potential from D-terms in equation (\ref{scalarV}), given by
\small{
\begin{eqnarray}
V_D&=&\frac{g_{1}^{2}}{8}[\lvert H_{d}^{-}\rvert^{2} +\lvert H_{d}^{0}\rvert^{2}-\lvert H_{u}^{+}\rvert^{2}-\lvert H_{u}^{0}\rvert^{2}  ]^2 \nonumber \\
&& + \frac{g_{2}^{2}}{8}[\lvert H_{d}^{-}\rvert^{2} +\lvert H_{d}^{0}\rvert^{2}-\lvert H_{u}^{+}\rvert^{2}-\lvert H_{u}^{0}\rvert^{2} +2\lvert
 T_0^+\rvert^2- 2\lvert T_0^-\rvert^2]^2  \nonumber \\ && + \frac{g_{2}^{2}}{8} [H_d^{0*}H_d^- + H_u^{+*} H_u^0 +\sqrt{2} (T_0^+ + T_0^-)T_0^* +h.c]^2
\nonumber \\ && -\frac{g_{2}^{2}}{8} [H_d^{-*}H_d^0 + H_u^{0*} H_u^+ +\sqrt{2} (T_0^+ - T_0^-)T_0^* +h.c]^2
\end{eqnarray}
}

\subsubsection{EWSB}
After Electroweak symmetry breaking, only the neutral components of the scalars fields  acquire vev's, i.e,
\begin{center}
 $\langle H_{u}^{0}\rangle=v_{u}$ , $\langle H_{d}^{0}\rangle=v_{d}$ , $ \langle S\rangle=v_{s}$ and $ \langle T^{0}\rangle=v_{t}$
\end{center}
The neutral-scalar part of the chiral superfields can be decomposed into real and imaginary parts,
\small{
\begin{eqnarray}
H_{u}^{0} &=& (H_{u_{R}}^{0} + v_{u}) + iH_{u_{I}}^{0}\\
H_{d}^{0} &=& (H_{d_{R}}^{0} + v_{d}) + iH_{d_{I}}^{0}\\
S &=& (S_R +v_s)+ iS_I\\
T^{0} &=& (T_{R}^{0} + v_{t}) + iT_{I}^{0}
\end{eqnarray}
}
The minimization conditions are derived from the fact that,
\begin{equation}
\frac{\partial V}{\partial v_u}=\frac{\partial V}{\partial v_d}=\frac{\partial V}{\partial v_s}=\frac{\partial V}{\partial v_t}=0
\end{equation}
We can determine the soft breaking mass parameters like $m_{H_u}^2$, $m_{H_d}^2$, $m_T^2$ and $ m_S^2 $ using the
following minimization conditions,
\small{
\begin{eqnarray}
m_{H_u}^2 &=& \cot\beta[A_{eff}-(\lambda^2+\frac{\lambda_2^2}{2})\frac{v^2}{2}\sin2\beta+\lambda\lambda_4v_t^2-\sqrt{2}\lambda_2\lambda_4v_tv_s
\nonumber \\ &&-\frac{\lambda_2}{\sqrt{2}}A_{\lambda_2}v_t]-\mu_{eff}^2+\frac{1}{4}(g_1^2+g_2^2)v^2 \cos2\beta\\
m_{H_d}^2 &=&\tan\beta[A_{eff}- (\lambda^2+\frac{\lambda_2^2}{2})\frac{v^2}{2}\sin2\beta+\lambda\lambda_4v_t^2-\sqrt{2}\lambda_2\lambda_4v_tv_s
\nonumber \\ &&-\frac{\lambda_2}{\sqrt{2}}A_{\lambda_2}v_t]-\mu_{eff}^2-\frac{1}{4}(g_1^2+g_2^2)v^2 \cos2\beta\\
m_S^2 &=&v^2[\frac{v_t}{\sqrt{2} v_s}\lambda \lambda_2+\lambda \lambda_1 \sin2\beta+\frac{1}{2v_s}\lambda A_\lambda \sin2\beta-\lambda^2]
 -[2\lambda_1^2 v_s+\lambda A_{\lambda_1}]v_s\nonumber \\ &&-\lambda_4v_t^2[B_{\lambda}/v_s+2\lambda_1+4\lambda_4]-\sqrt{2}\lambda_2\lambda_4v_uv_dv_t/v_s\\
m_T^2 &=&[\frac{1}{\sqrt{2}}\lambda \lambda_2\frac{v_s}{v_t}-\frac{\lambda_2^2}{2}-\frac{\lambda_2}{2\sqrt{2} v_t}A_{\lambda_2}\sin2\beta]v^2-2\lambda_4^2v_t^2+
2\lambda\lambda_4v_uv_d \nonumber \\ &&-\lambda_4v_s^2[2B_{\lambda}/v_s+2\lambda_1+4\lambda_4]-\sqrt{2}\lambda_2\lambda_4v_uv_dv_s/v_t
\end{eqnarray}
}
where, 
\begin{equation}
 A_{eff} = \lambda v_s[A_{\lambda}+\lambda_1v_s]
\label{aeff}
\end{equation}

and $v_{u}^{2}+v_{d}^{2} = v^{2}=(174)^{2} GeV^{2}$, $\tan\beta= \frac{v_{u}}{v_{d}}$.


Due to the addition of the triplets, the gauge bosons receive additional contribution in their masses like,
\begin{eqnarray}
 M_Z^2 &=&\frac{1}{2}(g_1^2+g_2^2)v^2\\
M_W^2 &=& \frac{1}{2}g_2^2(v^2+4 v_t^2)
\end{eqnarray}
The $\rho $-parameter at the tree-level is defined as,
\begin{equation}
 \rho=\frac{M_W^2}{M_Z^2 \cos^2 \theta_W}=1+4\frac{v_t^2}{v^2}
\end{equation}
Clearly, the $\rho$-parameter deviates from unity by a factor of $4\frac{v_t^2}{v^2}$. Using the recent bound
on $\rho$-parameter at 95$\%$ C.L. we can determine the bound on the triplet Higgs vev $v_t$. $\rho$ can be confined in the range
0.9799-1.0066 \cite{Espinosa:1991wt} and hence $v_t\leq 9$ GeV at  95$\%$ C.L.

\section{Higgs Sector}
\label{higgs}

\subsection{CP-even Higgs Mass Matrices}
The symmetric CP-even Higgs mass matrix is written in the basis of ( $H_{u_R}^0$ , $H_{d_R}^0$ , $T_R^0$ , $S_R$ ) with 10
independent components. After Electroweak symmetry breaking (EWSB) the entries of the squared mass-matrix are,
\small{
\begin{eqnarray}
 M^2_{11}&=&\frac{1}{2}(g_1^2+g_2^2)v^2\sin^2\beta+C_1\cot\beta+C_4,\nonumber \\
 M^2_{22}&=&\frac{1}{2}(g_1^2+g_2^2)v^2\cos^2\beta+C_1\tan\beta+C_4,\nonumber \\
 M^2_{33}&=&4\lambda_4^2v_t^2+\lambda_2 v^2[\lambda v_s-(A_{\lambda_2}+2\lambda_4v_s)\sin\beta\cos\beta]/\sqrt{2}v_t,\nonumber \\
 M^2_{44}&=&\lambda_1v_s[A_{\lambda_1}+4\lambda_1v_s]+[v_t(\lambda\lambda_2\frac{v^2}{\sqrt{2}}-\lambda_4B_\lambda v_t)
\nonumber \\ && +(\lambda A_\lambda-\sqrt{2}\lambda_2\lambda_4v_t)v^2\sin\beta\cos\beta]/v_s,\nonumber \\
 M^2_{12}&=&-C_1+[2\lambda^2+\lambda_2^2-\frac{(g_1^2+g_2^2)}{2}]v^2\sin\beta\cos\beta,\nonumber \\
 M^2_{13}&=&v[C_2\cos\beta -\sqrt{2} \lambda_2 \mu_{eff}\sin\beta],\nonumber \\
 M^2_{14}&=&-v[C_3\cos\beta -2 \lambda \mu_{eff}\sin\beta],\nonumber \\
 M^2_{23}&=&v[C_2\sin\beta -\sqrt{2} \lambda_2 \mu_{eff} \cos\beta],\nonumber \\
 M^2_{24}&=&-v[C_3\sin\beta -2\lambda \mu_{eff}\cos\beta],\nonumber \\
 M^2_{34}&=& 2\lambda_4v_t[B_\lambda+2v_s(\lambda_1+2 \lambda_4)]-\lambda_2 v^2(\lambda-2\lambda_4\sin\beta\cos\beta)/\sqrt{2}
\end{eqnarray}
}
where $C_i$'s are defined as,
\small{
\begin{eqnarray}
 C_1 &=& A_{eff}+\lambda\lambda_4v_t^2-\lambda_2A_{\lambda_2}\frac{v_t}{\sqrt{2}}-\sqrt{2}\lambda_2\lambda_4v_tv_s ,\nonumber\\
 C_2 &=& \frac{\lambda_2 A_{\lambda_2}}{\sqrt{2}}-2\lambda\lambda_4v_t+\sqrt{2}\lambda_4\lambda_2v_s,\nonumber \\
 C_3 &=& \lambda A_\lambda+2\lambda\lambda_1v_s-\sqrt{2}\lambda_2\lambda_4v_t,\nonumber\\
 C_4 &=& \lambda_2v_t[\frac{\lambda_2v_t}{2}-\sqrt{2}\lambda v_s]
\label{c4}
\end{eqnarray}
}
and $A_{eff}$ is defined in equation(\ref{aeff}).
\begin{itemize}
 \item \it Bound on the lightest Higgs mass :
\end{itemize}
 The bound on the lightest Higgs mass is derived from the fact that,
the smallest eigenvalue of a real, symmetric $n\times n$ matrix is smaller than the smallest eigenvalue of the upper
left $2\times2$ sub-matrix\cite{Espinosa:1991gr}. Using this we obtain an upper bound on the lightest CP-even Higgs mass,
\begin{eqnarray}
m_{h}^{2}\leqslant M_Z^2\left[\cos^2 2\beta+\frac{2\lambda ^2}{g_1^2+g_2^2}\sin^2 2\beta+\frac{\lambda_2^2}{g_1^2+g_2^2}\sin^2 2\beta \right]
\label{bound}
\end{eqnarray}
The bound on lightest Higgs mass has been considerably improved over the MSSM due to the
additional contribution from the singlet and triplet gauge fields. Using equation(\ref{bound})
we can put constraints on the parameters like $\lambda$, $\lambda_2$ and
$\tan\beta$ satisfying the recent bound on Higgs mass from ATLAS and CMS.

\subsection{CP-odd Higgs Mass Matrices}
The elements of the $4\times 4$ CP-odd Higgs squared mass matrix, after EWSB, in the basis of
( $H_{d_I}^0$ , $H_{u_I}^0$ , $S_I$ , $T_I^0$ ) are,
\small{
\begin{eqnarray}
 M^2_{P_{11}}&=&C_1 \tan\beta+C_4,\nonumber \\
 M^2_{P_{22}}&=&C_1 \cot\beta+C_4,\nonumber \\
 M^2_{P_{33}}&=&-3\lambda_1A_{\lambda_1}v_s-\lambda_4[B_\lambda+4\lambda_1v_s]\frac{v_t^2}{v_s}+D_1(\frac{v_t}{v_s})
+[\lambda A_\lambda/v_s+4\lambda\lambda_1 ]v^2\sin\beta\cos\beta,\nonumber\\
 M^2_{P_{44}}&=&-4\lambda_4v_s[B_\lambda+\lambda_1v_s]+D_1(\frac{v_s}{v_t})+[4\lambda\lambda_4-\frac{1}{\sqrt{2}v_t}\lambda_2A_{\lambda_2}
]v^2\sin\beta\cos\beta,\nonumber\\
 M^2_{P_{12}}&=&A_{eff}-\frac{v_t}{\sqrt{2}}\lambda_2 A_{\lambda_2}+\lambda_4v_t[\lambda v_t-\sqrt{2} \lambda_2v_s],\nonumber \\
 M^2_{P_{13}}&=&v\sin\beta[\lambda A_{\lambda}-2\lambda\lambda_1v_s+\sqrt{2} \lambda_2\lambda_4v_t] ,\nonumber \\
 M^2_{P_{14}}&=&-v\sin\beta[2\lambda\lambda_4v_t+\frac{1}{\sqrt{2}} \lambda_2(A_{\lambda_2}-2\lambda_4 v_s)],\nonumber \\
 M^2_{P_{23}}&=&M^2_{P_{13}}/\tan\beta,\nonumber \\
 M^2_{P_{24}}&=&M^2_{P_{14}}/\tan\beta,\nonumber \\
 M^2_{P_{34}}&=&-2\lambda_4v_t(B_\lambda-2\lambda_1v_s)-D_1 
\end{eqnarray}
}
where,
\begin{eqnarray}
 D_1 &=& \frac{1}{\sqrt{2}}\lambda_2 v^2(\lambda+2\lambda_4\sin\beta\cos\beta)
 \end{eqnarray}
This matrix always contains a Goldstone mode $G^0$ (gives mass to Z-boson), which can be written as,
\begin{equation}
 G^0 = \cos\beta H_{d_I}^0-\sin\beta H_{u_I}^0
\label{goldstone}
\end{equation}
and we rotate the mass matrix in the basis ( $G^0$, $A_1$, $A_2$, $A_3$ ) where,
\begin{equation}
 \begin{pmatrix}
 A_1 \\
 G^0 \\
 A_2 \\
 A_3
 \end{pmatrix}=\begin{pmatrix}
               \cos\beta &\sin\beta &0 &0\\
              -\sin\beta &\cos\beta &0 &0\\
               0 &0 &1 &0\\
               0 &0 &0 &1
               \end{pmatrix}
             \begin{pmatrix}
               H_{u_I}^0 \\
               H_{d_I}^0 \\
               S_I \\
               T_I^0
             \end{pmatrix}
\end{equation}

After removing the Goldstone mode, we again rotate the remaining
3$\times$3 mass matrix and finally obtain,
\small{
\begin{eqnarray}
 P_1&=& \cos\alpha \sin\beta H_{d_I}+\cos\alpha\cos\beta H_{u_I}+\sin\alpha S_I,\nonumber \\
P_1&=& -\sin\alpha \sin\beta H_{d_I}-\sin\alpha\cos\beta H_{u_I}+\cos\alpha S_I,\nonumber\\
P_3&=&T_I
\end{eqnarray}}
where $P_1$, $P_2$, $P_3$ are the massive modes.

\subsection{Charged Higgs Mass Matrices}

The charged Higgs sector comprises of a $4\times4$ symmetric matrix, written in the basis ($H_u^+$, $H_d^{-^*}$,
$T^{+}_{0}$, $T^{-^*}_{0}$), which has 10 independent components, (after EWSB) given by
\small{
\begin{eqnarray}
 (M^2_\pm)_{11} &=& E_1v_d^2+[\sqrt{2}\lambda\lambda_2v_tv_s+\frac{\lambda_2^2}{2}v_t^2]+C_1\cot\beta,\nonumber\\
 (M^2_\pm)_{12} &=& A_{eff}+E_1v_uv_d+[\lambda_2A_{\lambda_2}+\sqrt{2} \lambda v_t+2\lambda_2v_s]\frac{v_t}{\sqrt{2}},\nonumber\\
 (M^2_\pm)_{13} &=& E_2 v_d-2\lambda_2 v_u[\lambda v_s+\frac{\lambda_2v_t}{\sqrt{2}}],\nonumber\\
 (M^2_\pm)_{14} &=& E_3v_d+v_u\lambda_2\mu_{eff},\nonumber\\
 (M^2_\pm)_{22} &=& E_1v_u^2+[\sqrt{2}\lambda\lambda_2v_tv_s+\frac{\lambda_2^2}{2}v_t^2]+C_1\tan\beta,\nonumber\\
 (M^2_\pm)_{23} &=& E_3v_u+v_d\lambda_2\mu_{eff},\nonumber\\
 (M^2_\pm)_{24} &=& E_2v_u-2\lambda_2 v_d[\lambda v_s+\frac{\lambda_2v_t}{\sqrt{2}}] ,\nonumber\\
 (M^2_\pm)_{33} &=& \frac{g_2^2}{2}[v_u^2-v_d^2]+\lambda_2^2v_u^2+E_4,\nonumber\\
 (M^2_\pm)_{34} &=& [g_2^2-2\lambda_4^2]v_t^2-2\lambda_4v_s[B_\lambda+\lambda_1v_s]+2\lambda\lambda_4v_uv_d,\nonumber\\
 (M^2_\pm)_{44} &=& \frac{g_2^2}{2}[v_d^2-v_u^2]+\lambda_2^2v_d^2+E_4
\end{eqnarray}
}
where $E_i$'s are defined as,
\begin{eqnarray}
 E_1 &=& \frac{g_2^2}{2}-\lambda^2+\frac{\lambda_2^2}{2}, \nonumber\\
 E_2 &=& \frac{g_2^2v_t}{\sqrt{2}}+2\lambda_2\lambda_4v_s , \nonumber\\
 E_3 &=& \frac{g_2^2v_t}{\sqrt{2}}-\lambda_2A_{\lambda_2}, \nonumber\\
 E_4 &=& g_2^2v_t^2+4\lambda_4^2v_s^2                       
\end{eqnarray}

After diagonalization, we obtain one massless Goldstone state $G^+$ (gives mass to $W^\pm$-boson, since $G^-\equiv G^{+*}$),
\begin{equation}
 G^+ = \sin\beta H_u^+ - \cos\beta H_d^{-*} +\sqrt{2}\frac{v_t}{v}(T_0^+ - T_0^{-*})
\end{equation}
and three other massive modes like $H_1^\pm, H_2^\pm, H_3^\pm$


\section{Neutralinos and Charginos}
\label{neut}
The neutralino mass matrix extended by the singlet and triplet sector, in the basis
($\tilde{B},\tilde{W^{0}},\tilde{H_{d}^{0}},\tilde{H_{u}^{0}},\tilde{S},\tilde{T^0}$)
is given by,
\begin{equation}
\mathcal{M}_{\bar{G}} = \begin{pmatrix}
  M_{1}&  0&  -c_{\beta}s_{w}M_Z&  s_{\beta}s_{w}M_Z&  0&  0\\
  0&  M_{2}&  c_{\beta}c_{w}M_Z&  -s_{\beta}c_{w}M_Z&  0&  0\\
 -c_{\beta}s_{w}M_Z&  c_{\beta}c_{w}M_Z&  0& -\mu_{eff}&  -\lambda v_{u}& \frac{\lambda_{2}}{\sqrt{2}}v_u\\
  s_{\beta}s_{w}M_Z& -s_{\beta}c_{w}M_Z& -\mu_{eff}& 0& -\lambda v_{d}& \frac{\lambda_{2}}{\sqrt{2}}v_d\\
  0&  0& -\lambda v_{u}&  -\lambda v_{d}& 2\lambda_1 v_s& 2\lambda_4 v_t\\
  0&  0& \frac{\lambda_{2}}{\sqrt{2}}v_u&  \frac{\lambda_{2}}{\sqrt{2}}v_d& 2\lambda_4 v_t& 2\lambda_4 v_s
 \end{pmatrix}
\end{equation}
where, $M_1$, $M_2$ are the soft breaking mass parameters for Bino and Wino respectively and
\begin{center}
 $c_\beta = \cos\beta$, $s_\beta=\sin\beta$, $c_w=\cos\theta_w $ and $s_w=\sin\theta_w $
\end{center}
The left-most 4$\times$4 entries are exactly identical with that in MSSM, except the $\mu_{eff}$-term which
is defined in equation(\ref{mu}). As the triplet and the singlet fermion does not have any
interaction with the neutral gauginos the right-most 2$\times$2 entries are zero.\\
The chargino mass terms in the Lagrangian can be written as,
\begin{equation}
 -\frac{1}{2}[\tilde{G}^{+T}M_c^T.\tilde{G}^- +\tilde{G}^{-T}M_c.\tilde{G}^+]
\end{equation}
where, the basis $\tilde{G}^+$ and $\tilde{G}^-$ are specified as,
\begin{center}
$ \tilde{G}^+ = \begin{pmatrix}
                \tilde{W}^+\\
                \tilde{H_u}^+\\
                \tilde{T}^+
               \end{pmatrix}$ , $ \tilde{G}^- = \begin{pmatrix}
                \tilde{W}^-\\
                \tilde{H_d}^-\\
                \tilde{T}^-
               \end{pmatrix}$
\end{center}
and the chargino matrix in the gauge basis is given by,
\begin{equation}
 M_c = \begin{pmatrix}
        M_2& \frac{1}{\sqrt{2}}g_2 v_d& g_2v_t\\
        \frac{1}{\sqrt{2}}g_2 v_u& \lambda v_s+\frac{\lambda_2}{\sqrt{2}}v_t& -\lambda_2 v_d\\
        -g_2v_t& \lambda_2 v_u& 2\lambda_4 v_s
       \end{pmatrix}
\end{equation}


\section{Results and Discussions}
\label{results}

\begin{figure}[ht!]
\vspace*{10 mm}
\begin{center}
\begin{tabular}{cc}
\includegraphics[scale=0.9]{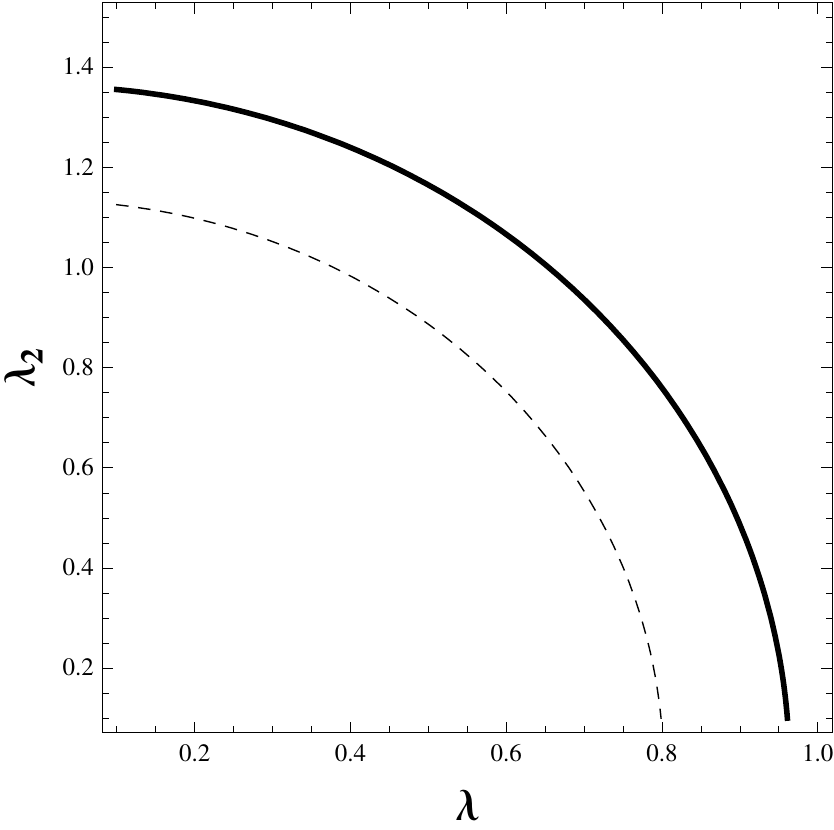}
\ &
\includegraphics[scale=0.9]{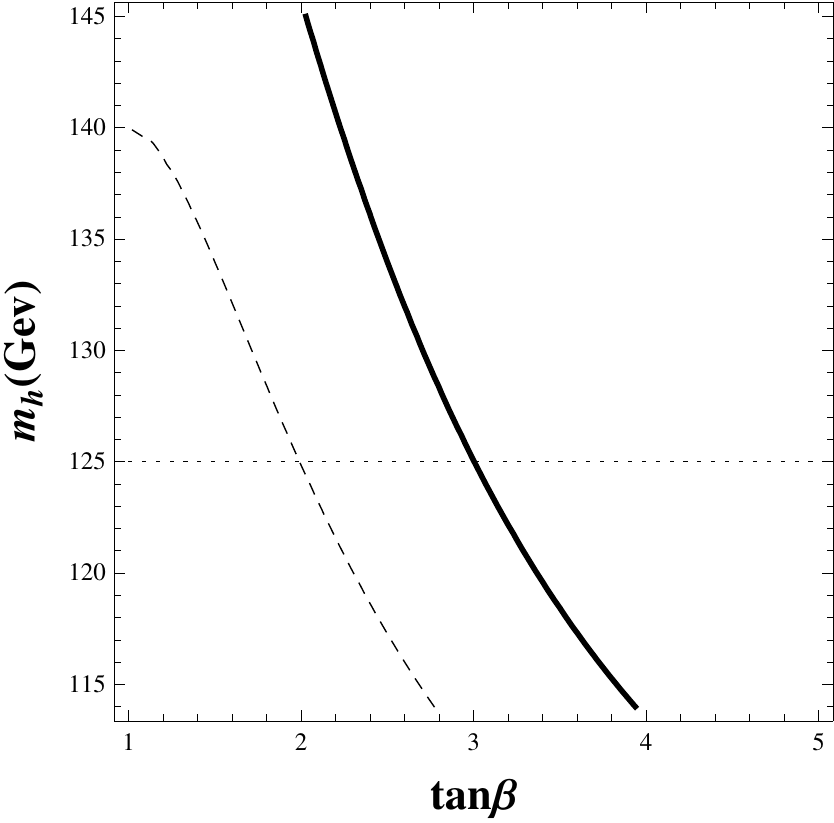}
\end{tabular}
\vspace*{3mm}
\caption{Left Panel : Plot of $\lambda$ vs. $ \lambda_2$, for $\tan\beta=2$ (dashed), 3 (thick) with $m_h=125$ GeV\\
Right Panel : Plot of $m_h$ vs. $\tan\beta$ for $\lambda = 0.6,\lambda_2= 0.75$ (dashed) $\lambda = 0.64,\lambda_2= 1.02$ (thick)
and the dotted line shows the recent bound i.e. $m_h=125$ GeV}
\label{fig:figure1}
\end{center}
\end{figure}

The main results of this paper are shown in Table(\ref{table 1}) and (\ref{table 2}). We have specified
the values of the parameters like couplings, soft-breaking parameters
at the Electroweak (EW) scale. The choice of $\tan\beta$, $\lambda$ and $\lambda_2$
are restricted from the bound on lightest Higgs mass(\ref{bound}).  In FIG.\ref{fig:figure1}
we show relation between $\lambda_2$ and $\lambda$ for different
values of $\tan\beta$. As we increase $\tan\beta$, $\lambda$ and $\lambda_2 $ tend to shift towards the higher
values.
 Plot in the right-hand panel of FIG.\ref{fig:figure1} shows the dependence
of $m_h$ on $\tan\beta$ for some particular choices of $\lambda=$0.6, 0.64 and $\lambda_2=$0.75, 1.02,
which are consistent with $m_h=125$ GeV (shown in the dotted line). In order to satisfy the bound
on Higgs mass, we can put constraint on $\tan\beta$
which is, $\tan\beta \leq 3.0$. The coupling $\lambda_1$ sets the mass for the singlino through the Yukawa term
$2 \lambda_1 S \chi_s\cdot \chi_s$. In order to have a light neutralino for satisfying the dark matter phenomenology we
choose  small values of $\lambda_1=0.2, 0.25$ as our benchmark values.The choice of $\lambda_4$ is determined from
the bounds on chargino masses. The other soft breaking parameters $A_\lambda, A_{\lambda_1},
A_{\lambda_2}, B_\lambda$ are chosen to fit the CP-even scalar masses specially to make the lightest Higgs mass close to 125 GeV. 
  Finally, we have chosen $\mu_{eff}$ to be $\mathcal O$(200 GeV)
and $v_t=2$ GeV, which determines the choice of $v_s$ from equation.(\ref{mu}).
The ratio  of $M_1$ to $M_2$ at the electroweak scale is
 consistent with universal gaugino masses at GUT scale and  gravity mediated SUSY breaking.

\begin{table}[t]
\caption{Value of the parameters specified at the Electroweak scale for two sets of Benchmark points.}
\vspace*{0.1cm}
\label{table 1}
 \begin{center}
 \begin{tabular}{|c|c|c|}\hline
Parameters at EW scale & Point 1 & Point 2\\\hline
$\tan\beta$ &2.0  &3.0  \\\hline
$\lambda$  & 0.60   &0.64\\\hline
$\lambda_1$  & 0.20   &0.25\\\hline
$\lambda_2$  & 0.75   &1.02\\\hline
$\lambda_4$  & 0.17   &0.20\\\hline
$\mu_{eff}$[GeV]  &200 &200\\\hline
$A_\lambda$[GeV] &400 &500 \\\hline
$A_{\lambda_1}$[GeV] &-10 &-10 \\\hline
$A_{\lambda_2}$[GeV] &600 &700 \\\hline
$B_\lambda$[GeV] &500 &600 \\\hline
$v_t$[GeV]  &2  &2 \\\hline
$M_1$[GeV]  &150 &200 \\\hline
$M_2$[GeV]  &300 &400 \\\hline
\end{tabular}
\end{center}
\end{table}

The mass spectrum shown in Table(\ref{table 2}) indicates all masses at the tree-level.
The Higgs spectrum consists of 4 CP-even Higgs (h, $H_1$, $H_2$, $H_3$),
3 pseudo-scalar Higgs ($A_1$, $A_2$, $A_3$) and 3 charged Higgs
($H_1^\pm$, $H_2^\pm$, $H_3^\pm$). We obtain significant contribution from the singlet and triplet sector
at the tree-level which is highly appreciable, since this has raised the mass of the
lightest CP-even Higgs boson to 125 GeV. Here we do not require a significant radiative  contribution
from the top-stop-sector \cite{King}. 
 The components of the lightest physical Higgs for $\tan\beta=2.0$ are given as,
\begin{equation}
 h=0.84205 H_{u_R}^0+ 0.44422 H_{d_R}^0+ 0.01977 T_R^0+ 0.30533 S_R
 \label{Higgs-frac}
\end{equation}
The lightest Higgs mass eigenstate has significant contribution from the Singlet and some contribution from the Triplet sectors.
We obtain the lightest scalar Higgs mass for the two sets of benchmark points as $120.6$ GeV and $119.2$ GeV respectively. This
will change the $h\rightarrow \gamma \gamma$ branching compared to the standard model and  precise determination of the Higgs 
decay branchings at LHC will be a good test of this model. In the pseudo-scalar Higgs sector, we obtain one Goldstone boson 
exactly identified as equation(\ref{goldstone}),i.e. $ G^0 = 0.4472 H_{d_I}^0-0.8942 H_{u_I}^0 $, for $\tan\beta=2.0$ and 

\begin{table}[!h]
\caption{Mass Spectrum and Relic Density for two sets of Benchmark points.}
\vspace*{0.1cm}
\label{table 2}
 \begin{center}
 \begin{tabular}{|c|c|c|}\hline
Mass Spectrum & Point 1 & Point 2\\\hline
\hline
\multicolumn{3}{|l|}{{\bf Neutral Higgs Spectrum}}
\\\hline
$m_h^{Tree}$[GeV]  &120.6 &119.2 \\\hline
$m_{H_1}$[GeV]  &145.5   &156.8 \\\hline
$m_{H_2}$[GeV]  &482.4   &630.7 \\\hline
$m_{H_3}$[GeV]  &825.2  &707.9 \\\hline
$m_{A_1}$[GeV]  &114.3   &116.9  \\\hline
$m_{A_2}$[GeV]  &487.8   &629.9  \\\hline
$m_{A_3}$[GeV]  &897.3  &816.0  \\\hline
\hline
\multicolumn{3}{|l|}{{\bf Charged Higgs Spectrum}}
\\\hline
$m_{H_1}^\pm$[GeV]  &208.4   &239.9 \\\hline
$m_{H_2}^\pm$[GeV]  &280.5   &320.6 \\\hline
$m_{H_3}^\pm$[GeV]  &496.3  &647.1 \\\hline
\hline
\multicolumn{3}{|l|}{{\bf Neutralino Spectrum}}
\\\hline
$m_{\tilde{\chi}_1^0}$[GeV]  &100.4   &102.9  \\\hline
$m_{\tilde{\chi}_2^0}$[GeV]  &122.6   &145.7  \\\hline
$m_{\tilde{\chi}_3^0}$[GeV]  &164.7   &205.9  \\\hline
$m_{\tilde{\chi}_4^0}$[GeV]  &212.6   &261.5  \\\hline
$m_{\tilde{\chi}_5^0}$[GeV]  &248.2   &265.7  \\\hline
$m_{\tilde{\chi}_6^0}$[GeV]  &345.0   &426.6  \\\hline
\hline
\multicolumn{3}{|l|}{{\bf Chargino Spectrum}}
\\\hline
$m_{\tilde{\chi}_1^{\pm}}$[GeV]  &124.2   &127.7  \\\hline
$m_{\tilde{\chi}_2^{\pm}}$[GeV]  &194.5   &250.2  \\\hline
$m_{\tilde{\chi}_3^{\pm}}$[GeV]  &347.1   &428.1  \\\hline
\hline
\multicolumn{3}{|l|}{{\bf Relic Density}}
\\\hline
$\Omega h^2$  &0.117   &0.08 \\\hline

\end{tabular}
\end{center}
\end{table}
\clearpage

$G^0 = 0.3163 H_{d_I}^0-0.9487 H_{u_I}^0 $, for $\tan\beta=3.0$. All other Higgs masses are listed in Higgs spectrum of 
Table(\ref{table 2}).\

The neutralino and the chargino sector consists of six and three mass-eigenstates respectively. The mass of the lightest neutralino
being $\cal{O}$(100 GeV), is the LSP of this model. The prospects of the LSP to be identified as a Dark Matter candidate is
discussed in detail in Section(\ref{DM}). Rest of the mass-spectrum are shown in Table(\ref{table 2})


\subsection{One-loop Correction to the Lightest Physical Higgs Mass}

The one-loop correction to $m_h^2$ is calculated by constructing the Coleman-Weinberg potential\cite{Coleman:1973jx},
\begin{equation}
 V_{CW}=\frac{1}{64\pi^2} STr[M^4(ln\frac{M^2}{Q_r^2}-\frac{3}{2})] 
\label{vcw}
\end{equation}
where $M^2$ are the field dependent tree-level mass matrices and $Q_r$ is the renormalization scale.
STr is the supertrace which includes a factor of $(-1)^{2J}(2J+1)$ and summed over the 
spin degrees of freedom. The one loop mass matrix can be derived from the above potential 
as follows,
\begin{eqnarray}
 (\Delta M_f^2)_{ij}&=& \frac{\partial^2 V_{CW}(f)}{\partial f_i \partial f_j}|_{vev}-
\frac{\delta_{ij}}{\langle f_i\rangle}\frac{\partial V_{CW}(f)}{\partial f_i}|_{vev}
\end{eqnarray}
where, $f_{i,j}$ stands for all the real components of $H_u^0, H_d^0$, S and $ T^0$.
Finally, the set of mass eigenvalues of the CP-even, CP-odd, Charged Higgs and Neutralino-Chargino 
mass matrices (all field-dependent) enters the calculation. The dominant contribution 
in the one-loop correction comes from the top-stop sector and 
the triplet sector. We compute the corrections only numerically using the benchmark values 
assigned for the sets of parameters. The results we obtain are given below in Table(\ref{table 3}) 
\begin{table}[h!]
\caption{Value of the lightest physical Higgs mass after 1-loop correction for two sets of Benchmark points.}
\vspace*{0.1cm}
\label{table 3}
 \begin{center}
 \begin{tabular}{|c|c|c|}\hline
Benchmark Point & $m_h^{Tree}$ [GeV] & $m_h^{Tree +Loop}$ [GeV] \\\hline
Point 1 &120.6  &124.9\\\hline
Point 2 &119.2  &125.5\\\hline
\end{tabular}
\end{center}
\end{table}

In both the cases we do not require large contribution from the radiative corrections to 
raise the lightest physical Higgs mass so as to satisfy the value of 125 GeV. 
This in turn implies that the contribution from the stop-top sector is not significant as in 
the case of MSSM. 
In fact in absence of fine tuning the correction to lightest physical Higgs mass from the 
stop-top sector is given by,
\begin{equation}
  \delta m_{H_u}^2(Q)\simeq \frac{3m_t^2}{(4\pi)^2v^2}ln\frac{m_{\tilde t_1}m_{\tilde t_2}}{m_t^2}
\end{equation}
For, $m_{\tilde t_1}$ and $m_{\tilde t_2}$ being $\cal O$ (200 GeV), this amounts to a correction of 
only a few GeV.


\subsection{Fine Tuning in the Electroweak Sector}

In this model, the lightest physical Higgs mass at the tree-level is boosted compared to 
NMSSM, and other triplet extended model \cite{DiChiara:2008rg} as it gets contribution from 
both singlet and triplet sector(\ref{bound}). Therefore, we can obtain a Higgs 
boson close to 125 GeV even at the tree level. After including the leading 
order radiative corrections from the stop-top and triplet sector, we get
\begin{equation}
 \delta m_{H_u}^2(Q)\simeq\frac{3y_t^2}{8\pi_2}(m_{\tilde t_1}^2+m_{\tilde t_2}^2+A_t^2)ln(\frac{Q}{M_z})+
\frac{3\lambda_2^2}{8\pi_2}(m_{T}^2+A_{\lambda_2}^2)ln(\frac{Q}{M_z})
\label{fine}
\end{equation} 
where, $m_{\tilde t_1}$ and $m_{\tilde t_2}$ are the soft masses of the stops, $A_t$ is the soft trilinear coupling, $y_t$ 
is the Yukawa coupling and Q is the fundamental scale of SUSY-breaking.\\
 The fine-tuning parameter can be quantified \cite{Barbieri,Ellwanger:2011mu}as,
\begin{equation}
 \Delta_{FT} \equiv \frac{m_{H_u}^2}{M_z^2} \frac{\partial M_z^2}{\partial m_{H_u}^2}
\end{equation}
 In case of MSSM (only first term in eqn.\ref{fine} is present), we have
\begin{eqnarray}
 \Delta_{FT}^{Stop} &\simeq& \frac{3y_t^2}{8\pi_2}(m_{\tilde t_1}^2+m_{\tilde t_2}^2+A_t^2)ln(\frac{Q}{M_z})
\end{eqnarray}
But the tree-level bound on Higgs mass is $m_h \leq M_z \cos 2\beta$. Therefore, one is forced to consider large 
values for $m_{\tilde t_1}$, $m_{\tilde t_2}$ and  $A_t$, say 1 TeV in order 
to raise the lightest physical Higgs Boson mass upto 125 GeV. In this case, $\Delta_{FT}^{Stop}\simeq 80$ 
and thus it leads to maximal stop mixing.\\ 
In NMSSM, the radiative correction needed to get a 125 GeV Higgs mass is $\delta m_h = 55$GeV. There is no doubt 
an improvement over MSSM, but still fine tuning is required in the stop-top sector\cite{King}.
In the model with one triplet \cite{DiChiara:2008rg}, the lightest physical Higgs mass can be raised 
to 113 GeV. Here, the required value of radiative correction is $\delta m_h = 53$ GeV. Now, 
the fine tuning due to the triplet sector is,
\begin{eqnarray} 
 \Delta_{FT}^{Trip} &\simeq & \frac{3\lambda_2^2}{8\pi_2}(m_{T}^2+A_{\lambda_2}^2)ln(\frac{Q}{M_z})
\end{eqnarray}
where $\lambda_2=0.8, 0.9$. The value of $\Delta_{FT}^{Trip}$ can be as large as 40. 
Therefore, this model can no longer be considered as a zero fine-tuning model.\\ 
Now coming to our model, we require $\delta m_h\simeq 35$ GeV only- here we see a distinct improvement of 20-50 GeV  
compared to other models discussed so far. Also, $\lambda_2=0.75, 1.02$ being comparable to $y_t$, 
we do not need heavy stops or large stop-top mixing to get the required Higgs mass. For example, $m_T=200$ GeV, 
$A_{\lambda_2}=700$ GeV and $Q=1$ TeV, we obtain $\Delta_{FT}^{Trip} \simeq 10$. Thus, we can 
achieve little fine-tuning compared to other models, since the lightest physical Higgs mass can be 
large at tree level and does not require large contribution from the radiative corrections. Here we note 
that, the Higgs-Triplet-Higgs coupling $\lambda_2$  (0.75 and 1.02) becomes non-perturbative at GUT scale. 
But, these choices of $\lambda_2$ actually helps to raise the Higgs mass close to 125 GeV at the tree-level. 
Other alternative could be of course having small $\lambda_2$, but then we would require large radiative 
corrections. Therefore, we improve the level of fine-tuning at the cost of giving up perturbativity of 
$\lambda_2$ at GUT scale.

\section{Dark Matter}
\label{DM}

We have analyzed the neutralino sector where the lightest neutralino (LSP),
is a mixture of Higgsino-Triplino and turns out to be a viable Dark Matter candidate. The components
of $\tilde{\chi_0}$ (for $\tan\beta=2.0$), i.e. the LSP are,
\begin{eqnarray}
 \tilde{\chi_0}&=&-0.321\tilde{B}+0.192\tilde{W_3^0}-0.323\tilde{H_d^0}+0.644\tilde{H_u^0}-0.213\tilde{S}+0.544\tilde{T^0}
\end{eqnarray}
Since the LSP has mass $\cal{O}$(100 GeV), there are two possibilities of
final states into which it can annihilate, i.e. (i)Fermion final states and (ii)Gauge Boson
final states. For annihilation into fermions, except $t\bar{t}$ it can go to any other
$f\bar{f}$ pairs via pseudo-scalar Higgs, Z-boson exchange and sfermion exchange.
But, if we consider the neutralino to be more like triplino, then its coupling with
Z-boson is forbidden. Generally, it can annihilate into gauge boson pairs via several processes
like chargino exchange, scalar Higgs exchange and Z-boson exchange.
But the dominant contribution comes from annihilation into $W^\pm$ via chargino exchange,
which finally leads to the Relic Density of 0.117, consistent with WMAP \cite{Wmap}.

The scalar interaction between the dark matter (i.e Neutralino LSP) and the quark is given by,
\begin{equation}
 {\cal{L}}_{scalar}=a_q \bar{\chi}\chi\bar{q}q
\end{equation}
where $a_q$ is the coupling between the quark and the Neutralino. The scalar cross section
for the Neutralino scattering off a target nucleus (one has to sum over the proton and neutrons
in the target) is given by,
\begin{equation}
 \sigma_{scalar}=\frac{4m_r^2}{\pi}(Zf_p+(A-Z)f_n)^2
\end{equation}
where, $m_r$ is the reduced mass of the nucleon and $f_{p,n}$ is the Neutralino coupling to
proton or neutron\cite{Jungman:1995df,Bertone:2004pz}, given by
\begin{equation}
\label{scalarterms}
f_{p,n} = \sum_{q=u,d,s}  f_{Tq}^{(p,n)} a_q  \frac{m_{p,n}}{m_q}  + \frac{2}{27}f_{TG}^{(p,n)}
\sum_{q=c,b,t} a_q \frac{m_{p,n}}{m_q},
\end{equation}
where $f_{Tu}^{(p)}=0.020 \pm 0.004, f_{Td}^{(p)}=0.026 \pm 0.005, f_{Ts}^{(p)}=0.118 \pm 0.062,
f_{Tu}^{(n)}=0.014 \pm 0.003, f_{Td}^{(n)}=0.036 \pm 0.008$ and $f_{Ts}^{(n)}=0.118 \pm 0.062$
\cite{fvalues}. $f_{TG}^{(p,n)}$ is related to these values by
\begin{equation}
f_{TG}^{(p,n)} = 1 - \sum_{q=u,d,s} f_{Tq}^{(p,n)}.
\end{equation}
The term in Eq.~\ref{scalarterms} which includes $f_{TG}^{(p,n)}$ results from the
coupling of the WIMP to gluons in the target nuclei through a heavy quark loop.

We can approximate $\frac{a_q}{m_q}\simeq \alpha/(s-m_h^2)$ where, $\alpha$
is the product of different coupling and mixings, $m_q$ is the mass of the quark  and $s=4m_\chi^2$ ($m_\chi$ being the Dark matter mass). The parameter $\alpha$
plays a crucial role in determining the spin-independent cross-section and is highly model
dependent. Using this we estimate  $\alpha \simeq 2\times 10^{-4} GeV^{-1}$ and the
value of the spin-independent cross-section is $10^{-43} cm^2$,
which is below the exclusion limits of XENON100\cite{Baudis:2012zs} and other direct detection experiments.


\section{Conclusions}
In this paper we have explored an extension of MSSM where the Higgs sector is extended by a singlet and a 
 $Y=0$ triplet superfield. This is the minimal model which gives a tree level Higgs mass of $\cal O$(119-120 GeV) 
and the one-loop correction can easily raise it to 125 GeV without significant contribution from 
the stop-top sector. However, $\lambda_2=0.75, 1.02$ (at Electroweak scale) becomes non-perturbative 
at the GUT scale, while all other couplings remain perturbative upto GUT scale - on the other hand 
this is the price we pay to retain small fine-tuning. 

In addition, we see that the triplino and singlino contributions to the neutralino mass 
matrix gives a viable dark matter candidate with mass around $100$ GeV 
which may be seen at the LHC from the missing transverse energy signals \cite{LHC-DM}. 
In MSSM and NMSSM the  problem for getting the correct relic density of dark matter is related 
to the necessity of choosing chargino and scalar masses to be in the multi TeV scale to fit the 
Higgs mass from radiative corrections. The DM mass in MSSM is around 700 GeV while in NMSSM it 
is possible to obtain viable DM in the 100 GeV range. The main advantage of our model for 
the dark matter is that since the sparticle masses need not be very large compared to the 
electroweak scale the 'WIMP miracle' is restored and we are able 
to get DM mass in the 100 GeV range over a large parameter space of our model.

The data from LHC with integrated luminosity of $5 fb^{-1}$ has not only given an indication the
Higgs mass but there is also a measurement of the Higgs decay branchings into different channels.
Detailed analysis \cite{Higgs-data} of the 125 GeV Higgs branching fractions seen at the LHC
indicates that the signal ratio for Higgs decay into two photons is larger than SM prediction
by a factor of $2.0 \pm 0.5$, decay into $WW^*$ and $ZZ^*$ channels is smaller than SM by a
factor of $0.5 \pm 0.3$ and into $bb $ and $\tau\tau$  channels it is factor $1.3 \pm 0.5$ consistent
with SM. The lightest CP-even Higgs (\ref{Higgs-frac}) has a sizable fraction of the singlet and the Higgs decay 
phenomenology will be distinguishable from the MSSM \cite{Carena} and 
likely to be similar to the NMSSM scenario \cite{MSSM2gamma, Ellwanger:2011aa}. 
But, there will be some contribution from the triplet sector too. The phenomenological aspects of the 
real-triplet extended SM has been studied in \cite{RealT2gamma}.  More data from LHC will pinpoint or 
rule out the extended Higgs sector models and it would be useful to study the singlet-triplet 
extended MSSM model in greater detail with emphasis on the LHC signal in the future.


\end{document}